\documentclass{aa}
\usepackage{graphics}

\def\la{\lower.5ex\hbox{$\; \buildrel < \over \sim \;$}}
\def\ga{\lower.5ex\hbox{$\; \buildrel > \over \sim \;$}}

\begin{document}      

   \title{Deep H{\sc i} observations\thanks{Based on 100-m Effelsberg radio telescope observations} of the surroundings of ram pressure stripped Virgo spiral galaxies}

   \subtitle{Where is the stripped gas?}

   \author{B.~Vollmer\inst{1} \& W.~Huchtmeier\inst{2}}

   \offprints{B.~Vollmer, e-mail: bvollmer@astro.u-strasbg.fr}

   \institute{CDS, Observatoire astronomique de Strasbourg, UMR 7550, 11, rue de l'universit\'e, 
     67000 Strasbourg, France \and
     Max-Planck-Institut f\"ur Radioastronomie, Auf dem H\"ugel 69, 53121 Bonn Germany 
   } 
          
   \date{Received / Accepted}

   \authorrunning{Vollmer \& Huchtmeier}
   \titlerunning{Deep H{\sc i} observations of stripped Virgo spirals}

\abstract{
Deep Effelsberg 100-m H{\sc i} observations of 5 H{\sc i} deficient Virgo spiral
galaxies are presented. No new extended H{\sc i} tail is found in these galaxies.
The already known H{\sc i} tail north of NGC~4388 does not significantly extend further
than a WSRT image has shown.   
Based on the absence of H{\sc i} tails in a sample of 6 Virgo spiral galaxies and a 
balance of previous
detections of extraplanar gas in the targeted galaxies we propose a global picture
where the outer gas disk (beyond the optical radius $R_{25}$) is {\rm removed} 
much earlier than expected by the classical ram pressure criterion. Based on the two-phase
nature of atomic hydrogen located in a galactic disk, we argue that the warm diffuse
H{\sc i} in the outer galactic disk is evaporated much more rapidly than the cold 
dense H{\sc i}. Therefore,
after a ram pressure stripping event we can only observe atomic hydrogen
which was cold and dense before it was removed from the galactic disk.
This global picture is consistent with all available observations.
We detect between 0.3\% and 20\% of the stripped mass assuming an initially
non-deficient galaxy and between  3\% and 70\% of the stripped mass assuming an
initially H{\sc i} deficient galaxy (def=0.4). Under the latter assumption we estimate
an evaporation rate by dividing the missing mass by the estimated time
to peak ram pressure from dynamical simulations. We find evaporation rates
between 10 and 100~M$_{\odot}$yr$^{-1}$.
}

\maketitle

\section{Introduction \label{sec:introduction}}

The H{\sc i} properties of cluster spiral galaxies are significantly different from those
of field spiral galaxies. Spiral galaxies located near the cluster center are often
H{\sc i} deficient (Chamaraux et al. 1980, Bothun et al. 1982, 
Giovanelli \& Haynes 1985, Gavazzi 1987, 1989) 
and their H{\sc i} disk sizes are considerably reduced (van Gorkom \& Kotanyi 1985, Warmels 1988, 
Cayatte et al. 1990, 1994). The observed small H{\sc i} disks together with unperturbed old 
stellar disks favor a ram pressure stripping scenario as the origin of the
H{\sc i} deficiency of Virgo cluster spirals.
Fourteen out of the 22 brightest Virgo spiral galaxies show an H{\sc i} deficiency
greater than 0.3 (Cayatte et al. 1994), i.e. they have lost more than half of their
initial reservoir of atomic hydrogen. This represents more than $10^9$~M$_{\odot}$ per 
galaxy. Despite the large missing mass, it is very difficult to observe 
this stripped gas. In the Virgo cluster an extended gas tail is only observed
in one spiral galaxy.

This exception is NGC~4388 which is located at a projected distance of 
1.3$^{\rm o}$ ($\sim 0.4$~Mpc\footnote{We use a distance of 17~Mpc to the Virgo cluster})
from the Virgo cluster center (M87) and hosts a Seyfert 2 nucleus.
Yoshida et al. (2002) discovered a very large H$\alpha$ plume that extends
up to $\sim$35~kpc north eastwards from the galactic disk which is seen almost edge-on. 
This region contains $\sim 10^{5}$~M$_{\odot}$ of ionized gas. Vollmer \& Huchtmeier (2003)
were the first who detected atomic hydrogen associated with this plume.
Oosterloo \& van Gorkom (2005) imaged this gas tail with the WSRT and found an
extent of $\sim 100$~kpc and a mass of $3.4 \times 10^8$~M$_{\odot}$.
This represents only about 20\% of the stripped gas assuming an H{\sc i}
deficiency of 0.8 (Cayatte et al. 1994).
 
These arguments only hold if the spiral galaxy had as much gas as a field galaxy
of the same optical diameter and the same morphology before the ram pressure
stripping event. On the other hand, cluster galaxies might have experienced other gas removing 
interactions before ram pressure stripping.
Possible interactions can be divided into two classes (for a review see
Gavazzi \& Boselli 2006): (i) the ``preprocessing''
of spiral galaxies through tidal interactions in infalling galaxy groups 
(Mihos 2004, Fujita 2004, Dressler 2004) and (ii) harassment (Moore et al. 1996, 1998),
viscous/turbulent stripping (Nulsen 1982) and/or thermal evaporation (Cowie \& McKee 1977)  
which occur once the galaxy resides within the cluster.
These mechanisms make the missing mass decreasing but not vanish.
We still expect several $10^8$~M$_{\odot}$ of stripped gas which is not detected. 

The question thus arises where the stripped gas is hidden and if we can or
should detect it. The underlying problem is the evolution of the multiphase ISM
once it is pushed out of the galactic disk by ram pressure.
Within the starforming disk the ISM is turbulent and consists of several phases:
the hot ionized ($\sim 10^6$~K), the warm neutral and ionized ($\sim 10^4$~K),
and the cold neutral ($\sim 100$~K) phase (see, e.g. Kulkarni \& Heiles 1988,
Spitzer 1990, McKee 1995). The neutral phase is not uniform but of fractal nature 
(Elmegreen \& Falgarone 1996). Braun (1997) analyzed the resolved neutral hydrogen emission 
properties of the 11 nearest spiral galaxies. He identified the high-brightness network of
H{\sc i} emission features with the cold neutral medium ($T \sim 100$~K)
and found that the bulk of atomic hydrogen is in this cold phase 
(between 60\% and 90\%). Braun (1997) also noted that the fractional 
line flux due to the cold phase drops
abruptly near the optical radius of a given galaxy. However, beyond this radius
a cool phase still exists in most of the galaxies, 
even though it represents only a few percent of the total H{\sc i} flux.

When the ISM is leaving the disk, its heating (kinematical heating by supernova
explosions and heating by stellar radiation) decreases and the whole gas
is surrounded by the hot intracluster medium (ICM).
At the same time ram pressure driven shocks propagate through the ISM.
Initially dense ISM regions might collapse and form stable globules (Vollmer et al. 2001)
or form stars, whereas tenuous regions might expand. The intracluster
medium does not only confine the stripped gas, but also causes evaporation (Cowie \& McKee 1977).
How efficient evaporation is depends on the geometry of the magnetic field
frozen into the ISM. A tangled field will increases the evaporation timescale
considerably (Cowie et al. 1981, Malyshkin \& Kulsrud 2001). 
Since the magnetic field geometry is not accessible, only direct observations
of the stripped gas at different wavelengths can help us to determine what happens
to the ISM once it has left the galactic disk.

In this article we approach this problem with deep single dish H{\sc i} observations
to search for atomic hydrogen far away from the galactic disks ($>20$~kpc) 
and a balance of previous detections of extraplanar gas. 
We selected 5 galaxies for which deep interferometric H{\sc i} observations
showed extraplanar gas. The observations are
described in Sect.~\ref{sec:observations} followed by the presentation of the
results (Sect.~\ref{sec:results}). These results are discussed in the framework
of the stripping of a two-phase atomic hydrogen taking into account existing detections of 
extraplanar gas (Sect.~\ref{sec:discussion}). Finally, we give our conclusions in 
Sect.~\ref{sec:conclusions}.

\section{Observations \label{sec:observations}}

In 2001--2003, we performed 21-cm line observations with the Effelsberg 100-m
telescope at different positions centered on
the systemic velocities of NGC~4402, NGC~4438, NGC~4501, and NGC~4522 with a bandwidth of 12.5~MHz. 
The two-channel receiver had a system noise of $\sim$30~K. The 1024 channel autocorrelator
was split into four banks with 256 channels each, yielding a
channel separation of $\sim$10~km\,s$^{-1}$. We further binned the channels to obtain
a final channel separation of $\sim$10~km\,s$^{-1}$ like the archival VLA data which
we use for comparison (NGC~4402, NGC~4501, NGC~4522). 
The galaxy's central position and four 
positions at a distance of one beam width (9.3$'$) to the NW, SW, SE, and
NE from the galaxy center were observed in on--off mode
(5~min on source, 5~min off source). In addition, we observed a sixth position
$6.5'$ west of the galaxy center for NGC~4438. 
\begin{table}
      \caption{Integration times and rms.}
         \label{tab:table}
      \[
         \begin{array}{lcccccc}
	   \hline
	   \hline
           \noalign{\smallskip}
	   {\rm \bf NGC~4402} & & & & & & \\
           \hline
           \noalign{\smallskip}
           {\rm position} & {\rm C} & {\rm NW} & {\rm W} & {\rm SW} & {\rm SE} & {\rm NE} \\
	   \noalign{\smallskip}
	   \hline
	   \noalign{\smallskip}
	   $$\Delta t$$\ {\rm (min)} & 120 & 120 & - & 120 & 120 & 120 \\ 
           \noalign{\smallskip}
	   \hline
	   \noalign{\smallskip}
	   {\rm rms\ (mJy)} & 2.6 & 1.8 & - & 2.0 & 1.5 & 3.5 \\
	   \noalign{\smallskip}
           \hline
	   \hline
           \noalign{\smallskip}
	   {\rm \bf NGC~4438} & & & & & & \\
           \hline
           \noalign{\smallskip}
           {\rm position} & {\rm C} & {\rm NW} & {\rm W} & {\rm SW} & {\rm SE} & {\rm NE} \\
	   \noalign{\smallskip}
	   \hline
	   \noalign{\smallskip}
	   $$\Delta t$$\ {\rm (min)} & 120 & 120 & 120 & 120 & 120 & 120 \\ 
           \noalign{\smallskip}
	   \hline
	   \noalign{\smallskip}
	   {\rm rms\ (mJy)} & 1.8 & 1.7 & 1.8 & 2.2 & 2.2 & 2.2 \\
	   \noalign{\smallskip}
           \hline
	   \hline
           \noalign{\smallskip}
	   {\rm \bf NGC~4501} & & & & & & \\
           \hline
           \noalign{\smallskip}
           {\rm position} & {\rm C} & {\rm NW} & {\rm W} & {\rm SW} & {\rm SE} & {\rm NE} \\
	   \noalign{\smallskip}
	   \hline
	   \noalign{\smallskip}
	   $$\Delta t$$\ {\rm (min)} & 120 & 120 & - & 120 & 120 & 120 \\ 
           \noalign{\smallskip}
	   \hline
	   \noalign{\smallskip}
	   {\rm rms\ (mJy)} & 1.8 & 1.0 & - & 1.4 & 1.6 & 1.7 \\
	   \noalign{\smallskip}
           \hline
	   \hline
           \noalign{\smallskip}
	   {\rm \bf NGC~4522} & & & & & & \\
           \hline
           \noalign{\smallskip}
           {\rm position} & {\rm C} & {\rm NW} & {\rm W} & {\rm SW} & {\rm SE} & {\rm NE} \\
	   \noalign{\smallskip}
	   \hline
	   \noalign{\smallskip}
	   $$\Delta t$$\ {\rm (min)} & 120 & 120 & - & 120 & 120 & 120 \\ 
           \noalign{\smallskip}
	   \hline
	   \noalign{\smallskip}
	   {\rm rms\ (mJy)} & 1.2 & 1.8 & - & 1.9 & 1.1 & 1.2 \\
	   \noalign{\smallskip}
           \hline
	   \hline
        \end{array}
      \]
\end{table}
Care was taken to avoid other Virgo cluster galaxies with velocities within our bandwidth
in all observations. We used 3C286 for pointing and flux calibration. 
The observation time was 120~min per position.
The resulting noise (Table~\ref{tab:table}) is partly determined by small amplitude 
interferences, but it is close to the theoretical noise of 2~mJy per hour 
of integration: on average $1\sigma=1.5$~mJy (varying from 1.1 to 2.0~mJy).
 
In addition we observed a field of about $1^{\circ} \times 1^{\circ}$ centered on the H{\sc i}
plume of NGC~4388.
The noise level of these spectra is largely determined by the closeness of M87 which has
a flux density of $\sim 220$~Jy at 1.4~GHz. The noise of our spectra varies between 
$1\sigma=2$ and $7$~mJy.

In order to compare our Effelsberg H{\sc i} spectra to interferometric data where the
galaxy is spatially resolved, we use VLA 21~cm data (Crowl et al. 2005, Vollmer et al.
in prep., Kenney et al. 2004). 
These data have spatial resolutions of $\sim 20''$ and channel separations
of 10~km\,s$^{-1}$. We clipped the data cubes at a level of 3~mJy/beam
and produced a synthesized single dish spectrum using a Gaussian beam of $9.3'$ HPBW.

\section{Results \label{sec:results}}

\subsection{NGC~4388}

\begin{figure*}
  \resizebox{\hsize}{!}{\includegraphics{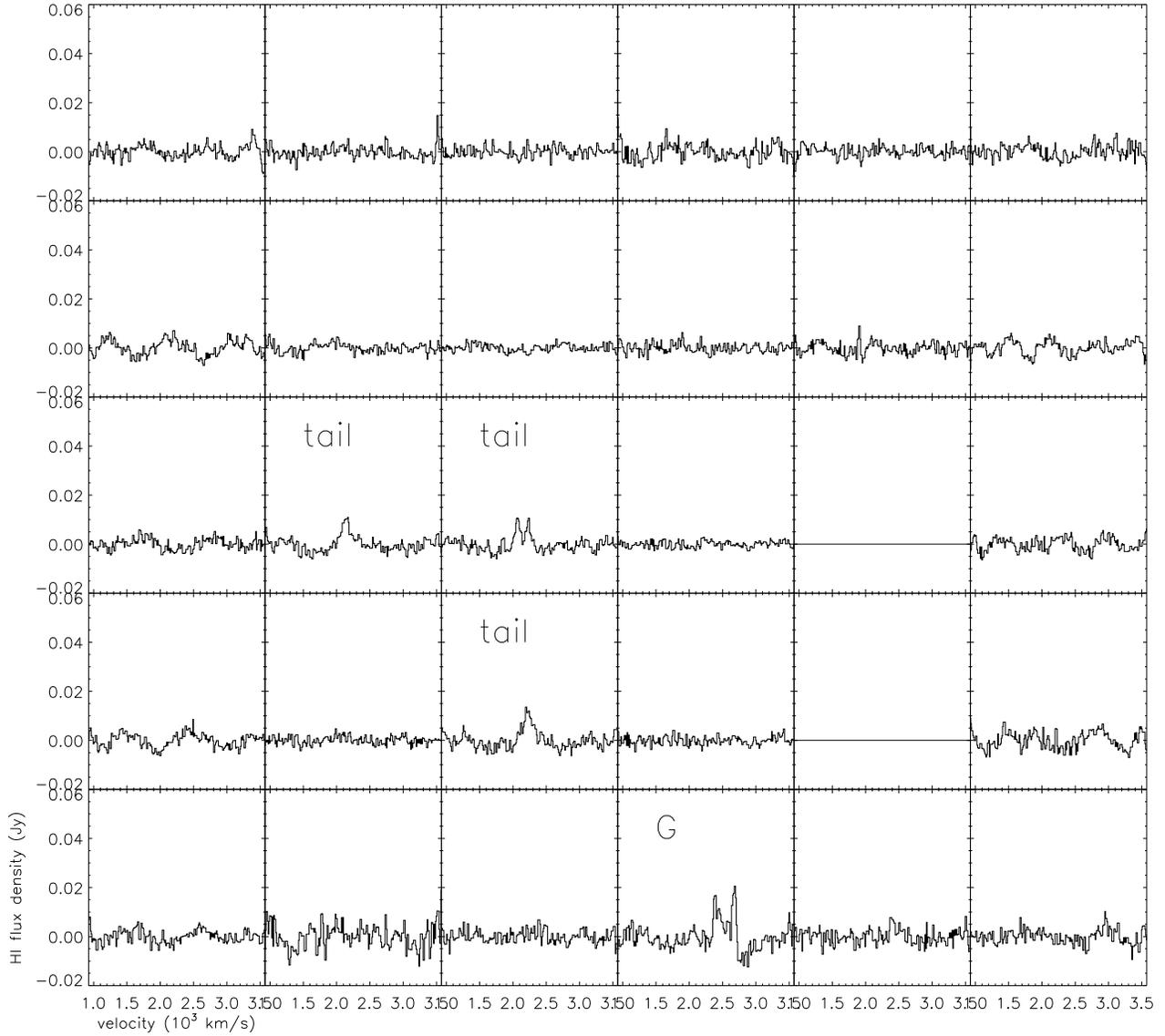}}
  \caption{Deep Effelsberg 100-m H{\sc i} spectra of the NGC~4388 H{\sc i} plume.
          The spectrum centered on the galaxy is labeled with a ``G''.
	  We clearly detect H{\sc i} line emission is 4 positions: in NGC~4388 and in 3 positions
	  to the north west of the galactic disk. These positions are labeled with ``tail''.
	  The spatial separation between two spectra corresponds to the
	  beamsize ($9.3'$). We were not able to obtain proper spectra for two 
	  positions north west of the galaxy.
	  In the corresponding boxes we have replaced the spectra by a solid line.
        } \label{fig:n4388ext}
\end{figure*}

We observed $6 \times 5$ positions centered on the H{\sc i} plume of NGC~4388 (Oosterloo \&
van Gorkom 2005) (Fig.~\ref{fig:n4388ext}). The spectrum centered on NGC~4388 is labeled
with a ``G''. The observing conditions did not permit us
to obtain interference-free spectra of two positions in the north-east of NGC~4388.
In the corresponding boxes we have replaced the spectra by a solid line.
The sinusoidal behaviour of the spectra in the west and the east is most probably due to
sidelobe detection of M87.
We clearly detect H{\sc i} line emission in 4 positions: in NGC~4388 and in 3 positions
to the north west of the galactic disk. These positions are labeled with ``tail''.
One of the spectra shows a double line profile consistent with the WSRT observations
of Oosterloo \& van Gorkom (2005). Our deep H{\sc i} observations show that there is
no significant amount of atomic hydrogen beyond this H{\sc i} tail.

\subsection{NGC~4402}

\begin{figure}
  \resizebox{\hsize}{!}{\includegraphics{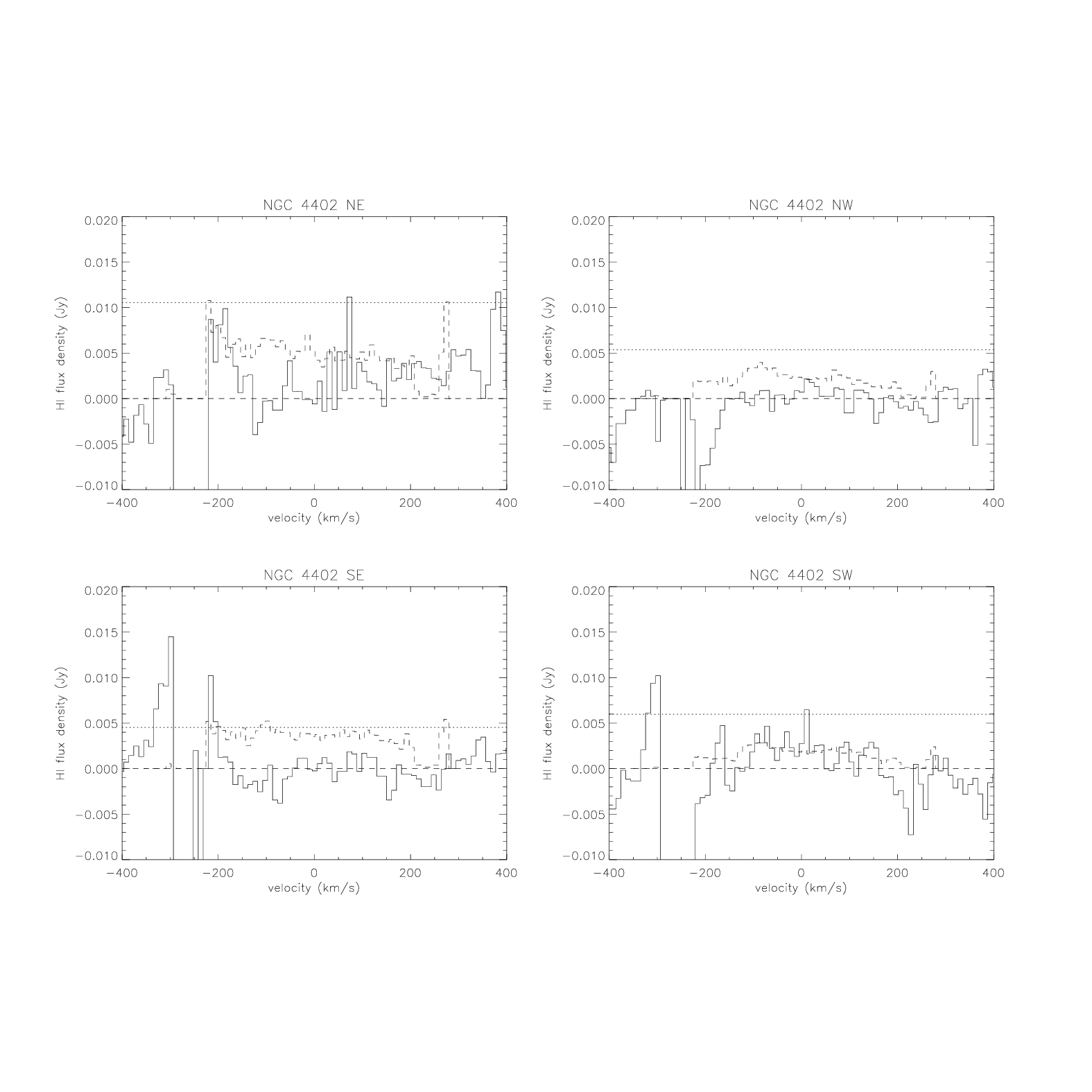}}
  \caption{Solid lines: Effelsberg 100-m spectra of the four off-center positions.
	Their locations with respect to the galaxy center are marked on top of each panel.
	Dashed line: synthesized VLA spectra (Crowl et al. 2005), which only show H{\sc i} disk emission. 
	Dotted line: 3$\sigma$ noise levels of the Effelsberg spectra. 
	Radial velocities are given relative to the systemic velocity of NGC~4402.
        } \label{fig:n4402eff}
\end{figure}

NGC~4402 is another H{\sc i} deficient edge-on spiral located close to NGC~4388.
Crowl et al. (2005) detected an asymmetric distribution of the 20cm continuum
emission and $2.7 \times 10^7$~M$_{\odot}$ of extraplanar atomic
hydrogen in the north east of the galactic disk.
Both features are consistent with a scenario where ram pressure is responsible for
the compressed radio continuum halo and the extraplanar H{\sc i}.
\begin{figure}
  \resizebox{\hsize}{!}{\includegraphics{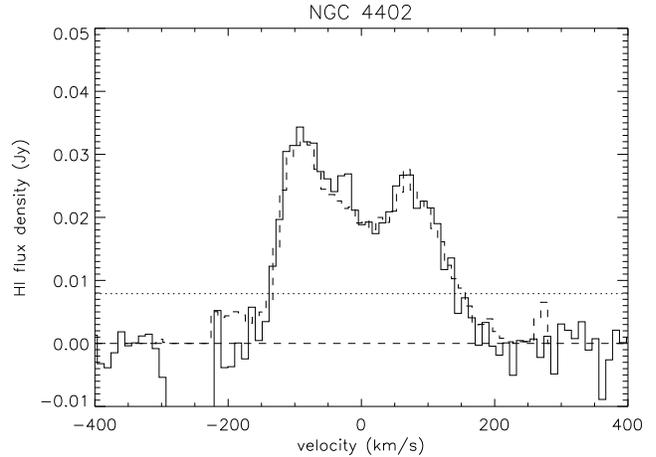}}
  \caption{Solid line: Effelsberg 100-m spectrum of the central position.
	Dashed line: spectrum of the VLA data (Crowl et al. 2005). Dotted line: 3$\sigma$ noise level of the
	100-m spectrum. Heliocentric velocities are given relative to the systemic velocity
	of NGC~4402 ($v_{\rm sys}$=235~km\,s$^{-1}$).
        } \label{fig:n4402c}
\end{figure}
Our Effelsberg observations of the central position Fig.~\ref{fig:n4402c}
do not reveal more H{\sc i} than observed with the VLA (Crowl et al. 2005). We do not find
any H{\sc i} in the offset positions (Fig.~\ref{fig:n4402eff}).
The gap in the spectrum around zero radial velocity is due to galactic H{\sc i} emission.

\subsection{NGC~4438}

NGC~4438 has a strongly perturbed stellar disk with a prominent stellar tidal arm
pointing to the north. This perturbation is due to a rapid and close gravitational
interaction with its companion S0 galaxy NGC~4435 (Combes et al. 1988, Vollmer et al. 2005).
The gas disk is heavily truncated and almost exclusively molecular.
Extraplanar CO with a mass of $\sim 5 \times 10^8$~M$_{\odot}$ is detected to the west of 
the galactic disk. Detailed modelling
of the interaction including the gravitational interaction, ram pressure, and
an ISM-ISM collision showed that ram pressure (together with the tidal
interaction) is the most important ingredient
to reproduce the observed CO emission distribution and kinematics (Vollmer et al. 2005).

\begin{figure}
  \resizebox{\hsize}{!}{\includegraphics{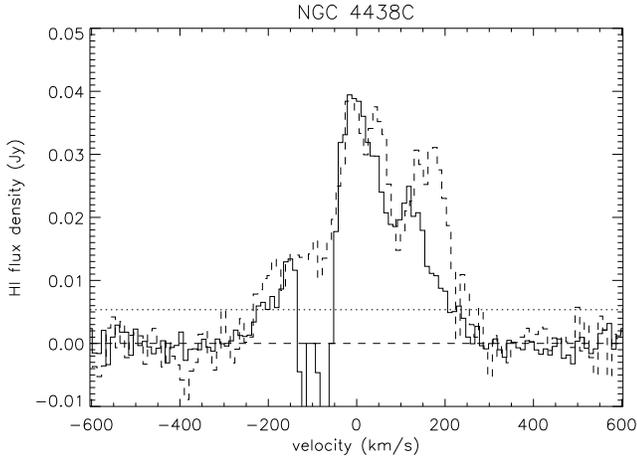}}
  \caption{Solid line: Effelsberg 100-m spectrum of the central position.
	Dashed line:  CO(1--0) spectrum from Vollmer et al. (2005) in arbitrary units.
	Dotted line: 3$\sigma$ noise level of the
	100-m spectrum. Heliocentric velocities are given relative to the systemic velocity
	of NGC~4438 ($v_{\rm sys}$=-70~km\,s$^{-1}$).
        } \label{fig:n4438c}
\end{figure}

\begin{figure}
  \resizebox{\hsize}{!}{\includegraphics{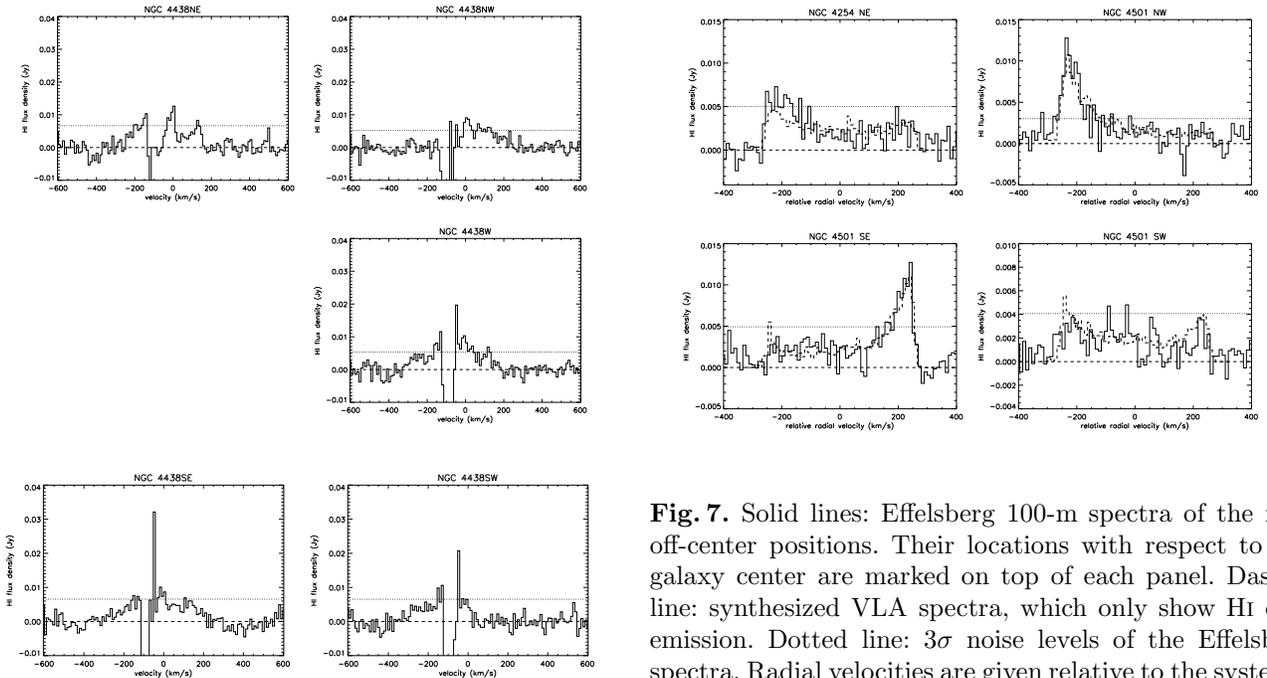}}
  \caption{Solid lines: Effelsberg 100-m spectra of the five off-center positions.
	Their locations with respect to the galaxy center are marked on top of each panel. 
	Dotted line: 3$\sigma$ noise levels of the Effelsberg spectra. 
	Radial velocities are given relative to the systemic velocity of NGC~4438.
        } \label{fig:n4438eff}
\end{figure}

With our deep Effelsberg observations, linearly interpolated where galactic
H{\sc i} emission dominates, we detect a total H{\sc i} mass of
$\sim 6 \times 10^8$~M$_{\odot}$ (Fig.~\ref{fig:n4438c}).
Since there is only a small fraction of the total H{\sc i} emission detected in 
interferometric radio observations
(Cayatte et al. 1990, Hibbard et al. 2001), we compare our deep Effelsberg H{\sc i}
spectrum with the integrated CO spectrum of Vollmer et al. (2005) where the
extraplanar gas is included. We find a remarkable resemblance between the
two spectra tracing the molecular and the atomic hydrogen.
The only differences are that (i) the ratio between the H{\sc i} peak flux density 
at $v=0$~km\,s$^{-1}$ and $v>0$~km\,s$^{-1}$ is larger than that of the
CO data and (ii) the CO peak at $v>0$~km\,s$^{-1}$ extends further to high velocities.
This might indicate that both gas phases are well mixed. However, without
deep interferometric H{\sc i} observations, which are still lacking
due to the closeness of M87, it is not possible to investigate the
H{\sc i} -- H$_2$ connection in more detail.
We do not find any significant H{\sc i} in the offset positions (Fig.~\ref{fig:n4438eff}).

\subsection{NGC~4501}

\begin{figure}
  \resizebox{\hsize}{!}{\includegraphics{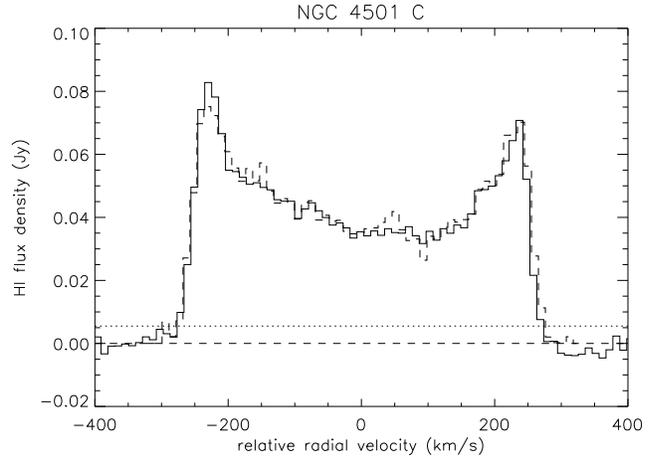}}
  \caption{Solid line: Effelsberg 100-m spectrum of the central position.
	Dashed line: spectrum of the VLA data (Vollmer et al. in prep.). 
	Dotted line: 3$\sigma$ noise level of the
	100-m spectrum. Heliocentric velocities are given relative to the systemic velocity
	of NGC~4501 ($v_{\rm sys}$=2281~km\,s$^{-1}$).
        } \label{fig:n4501c}
\end{figure}

\begin{figure}
  \resizebox{\hsize}{!}{\includegraphics{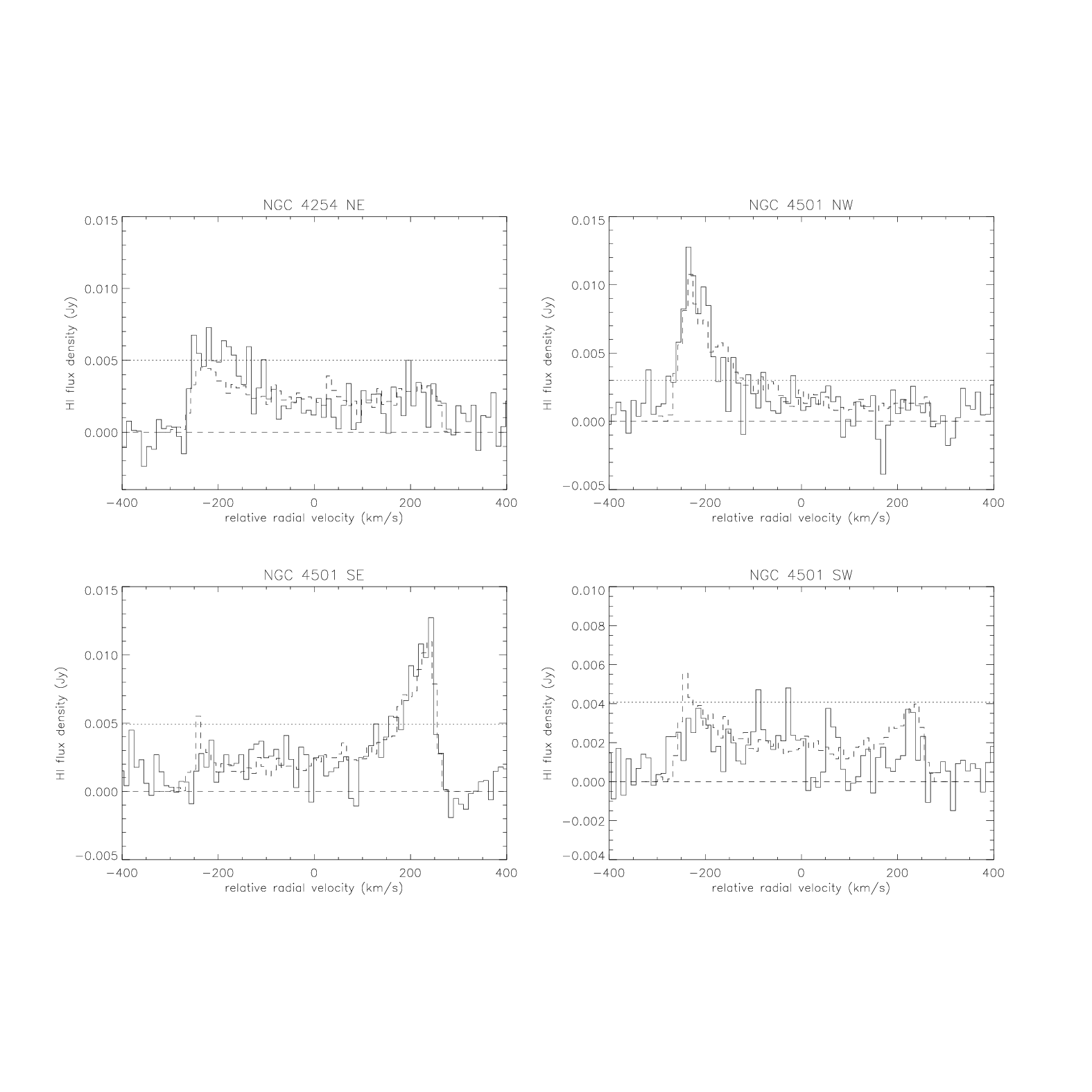}}
  \caption{Solid lines: Effelsberg 100-m spectra of the four off-center positions.
	Their locations with respect to the galaxy center are marked on top of each panel.
	Dashed line: synthesized VLA spectra, which only show H{\sc i} disk emission. 
	Dotted line: 3$\sigma$ noise levels of the Effelsberg spectra. 
	Radial velocities are given relative to the systemic velocity of NGC~4501.
        } \label{fig:n4501eff}
\end{figure}

The spiral galaxy NGC~4501 has an H{\sc i} disk which is truncated close to the
optical radius $R_{25}$. In addition, the H{\sc i} surface density is
enhanced in the south western side of the galactic disk (Cayatte et al. 1990, 1994).
In a forthcoming paper (Vollmer et al., in prep.) we will show that 
this galaxy is in a pre-peak stripping stage, i.e. it is approaching the
cluster center and ram pressure will reach its maximum in about 100~Myr.
Our deep Effelsberg H{\sc i} observations of the central position (Fig.~\ref{fig:n4501c})
do not reveal more H{\sc i} than observed with the VLA (Vollmer et al., in prep.). 
The single dish Effelsberg spectra of the offset positions show a possible
additional H{\sc i} flux density compared to the VLA data in the north-east
of the galactic disk (Fig.~\ref{fig:n4501eff}). This extra emission
corresponds to an H{\sc i} mass of $\sim 10^7$~M$_{\odot}$. 
This is consistent with the pre-peak stripping scenario.

\subsection{NGC~4522}

\begin{figure}
  \resizebox{\hsize}{!}{\includegraphics{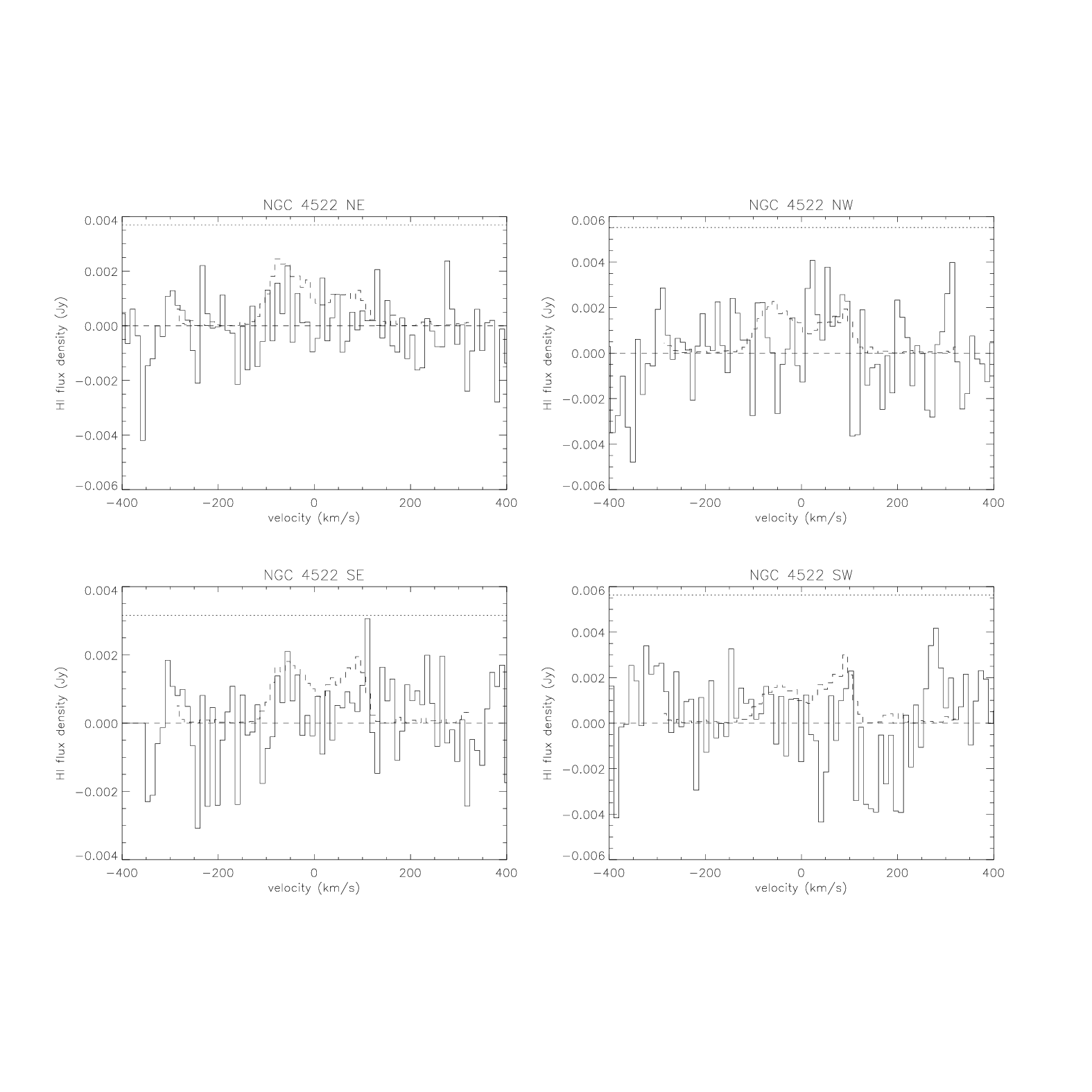}}
  \caption{Solid lines: Effelsberg 100-m spectra of the four off-center positions.
	Their locations with respect to the galaxy center are marked on top of each panel.
	Dashed line: synthesized VLA spectra (Kenney et al. 2004), which only show H{\sc i} disk emission. 
	Dotted line: 3$\sigma$ noise levels of the Effelsberg spectra. 
	Radial velocities are given relative to the systemic velocity of NGC~4522.
        } \label{fig:n4522eff}
\end{figure}

NGC~4522 is one of the best examples for ongoing ram pressure stripping. 
H{\sc i} and H$\alpha$ observations (Kenney et al. 2004, Kenney \& Koopmann 1999) showed a
heavily truncated gas disk at a radius of 3~kpc, which is $\sim 40$\% of the optical radius,
and a significant amount of extraplanar gas to the west 
of the galactic disk ($1.5 \times 10^8$~M$_{\odot}$). 
The one-sided extraplanar atomic gas distribution shows high column
densities, comparable to those of the adjectant galactic disk.
A strong ram pressure scenario can account for the truncated gas disk and the western 
extraplanar gas. Further evidence for such a peak ram pressure
scenario comes from polarized radio continuum observations (Vollmer et al. 2004b).
The 6~cm polarized emission is located at the eastern edge of the galactic disk, opposite to the 
western extraplanar gas. This ridge of polarized radio continuum emission is most likely
due to ram pressure compression of the interstellar medium (ISM) and its magnetic field. 
In addition, the degree of polarization decreases from the east to the west and the flattest 
spectral index between 20~cm and 6~cm coincides with the peak of the 6~cm polarized emission. 
These findings together with a detailed dynamical model (Vollmer et al. 2006) 
are consistent with a scenario where ram pressure is close to its maximum.
\begin{figure}
  \resizebox{\hsize}{!}{\includegraphics{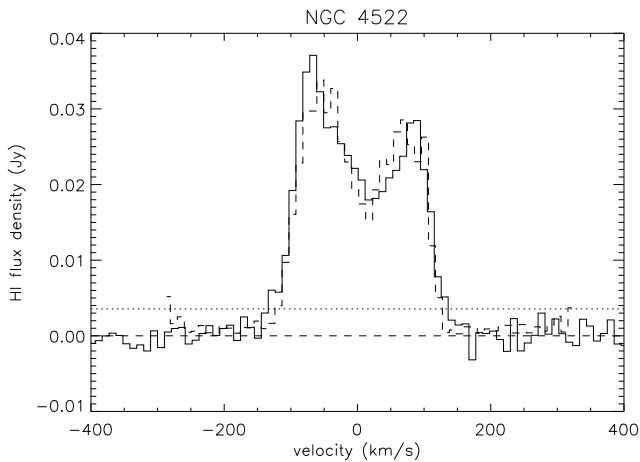}}
  \caption{Solid line: Effelsberg 100-m spectrum of the central position.
	Dashed line: spectrum of the VLA data of Kenney et al. (2004). Dotted line: 3$\sigma$ 
	noise level of the 100-m spectrum. 
	Heliocentric velocities are given relative to the systemic velocity
	of NGC~4522 ($v_{\rm sys}$=2324~km\,s$^{-1}$).
        } \label{fig:n4522c}
\end{figure}
Our Effelsberg observations of the central position (Fig.~\ref{fig:n4522c})
do not reveal more H{\sc i} than observed with the VLA (Kenney et al. 2004). We do not find
any H{\sc i} in the offset positions (Fig.~\ref{fig:n4522eff}).

\subsection{NGC~4569}

Deep VLA and Effelsberg H{\sc i} data together with a dynamical model of this galaxy are
presented in Vollmer et al. (2004a). They discovered a low surface density H{\sc i} arm
in the west of the galaxy, whose velocity field is distinct from that of the overall disk rotation.
No H{\sc i} emission was detected in the Effelsberg H{\sc i} off center observations.
A post--stripping scenario is consistent with the main observed characteristics of NGC~4569.
In this scenario the galaxy's closest approach to the cluster center, i.e. peak ram 
pressure, occured $\sim 300$~Myr ago.

\section{Where did the ram pressure stripped gas go? \label{sec:discussion}}

With our sample of 6 galaxies we can now investigate how much of the missing stripped
gas mass we detect in H{\sc i}. 
The first result is that we do not detect a significant 
amount of atomic hydrogen at distances greater than 20~kpc in any of these galaxies
except in NGC~4388 (see Sec.~\ref{sec:n4388}). 
For the determination of a gas detection rate we need to know
the expected atomic gas mass of a non-H{\sc i} deficient galaxy.
With the H{\sc i} deficiency and the total observed H{\sc i}
mass we can estimate the expected initial H{\sc i} mass of a spiral galaxy before it entered
the Virgo cluster (table~\ref{tab:masses}).
As discussed in Sec.~\ref{sec:introduction} this depends on whether the galaxy
experienced a gas removing interaction before the ram pressure stripping event.
Therefore we calculate the initial H{\sc i} mass for a ``normal'' field galaxy and
a galaxy that has already lost a significant amount of gas due to tidal
interactions, turbulent/viscous stripping, and/or evaporation.

\subsection{Initially ``normal'' gas rich spiral galaxies}

In general, the uncertainty of H{\sc i} deficiencies is about $\pm 0.2$.
We use the H{\sc i} deficiencies of Crowl et al. (2005) and Kenney et al. (2004)
which are consistent with those of Cayatte et al. (1994) within the errors.
If we assume that the missing (non-detected) gas has been evaporated by the
hot intracluster medium, we can estimate the evaporation rate by dividing
the missing mass by a characteristic timescale. 
For this timescale we chose the time to ram pressure peak when most of
the gas leaves the galaxy.
Detailed comparison between observations and numerical modelling of 
these galaxies showed that we observe NGC~4388 $\sim 100$~Myr after peak ram pressure
(Vollmer \& Huchtmeier 2003), NGC~4402 more than several 100~Myr after peak ram pressure
(Crowl et al. 2005), NGC~4438 is near peak ram pressure (Vollmer et al. 2005), 
NGC~4501 is $\sim 100$~Myr before peak ram pressure (Vollmer et al. in prep.), NGC~4522 is near
peak ram pressure (Vollmer et al. 2006), and NGC~4569 $\sim 300$~Myr after peak ram pressure
(Vollmer et al. 2004a).

Table~\ref{tab:masses} summarizes the data for all galaxies: the time to peak ram pressure (col.(2)),
the H{\sc i} deficiency (col.(3)),
the observed total H{\sc i} mass (col(4)), the observed extraplanar H{\sc i} gas mass
(col.(5); except for NGC~4438 where we have taken the extraplanar CO mass), the
expected H{\sc i} mass based on the H{\sc i} deficiency (col.(6)), the percentage of
extraplanar gas to the missing gas mass, i.e. the difference between the observed 
and the expected initial gas mass (col.(7)), the expected H{\sc i} mass assuming an initial
H{\sc i} deficiency of 0.4 (col.(8)), the percentage of
extraplanar gas to the missing gas mass assuming an initial H{\sc i} deficiency of 0.4 (col.(9)),
and the estimated evaporation rate (\ref{eq:evap}; col.(10)).
\begin{table*}
      \caption{Galaxy gas properties.}
         \label{tab:masses}
      \[
         \begin{array}{|l|c|c|c|c|c|c|c|c|c|}
            \hline
             {\rm name} & {\rm time\ to\ ram} & {\rm HI\ def} & {\rm M}^{\rm extra}_{\rm HI} &  {\rm M}^{\rm total}_{\rm HI} & {\rm M}^{\rm expected}_{\rm HI} & \%{\rm \ of} & {\rm M}^{\rm expected}_{\rm HI} & \%{\rm \ of} & {\rm evaporation} \\ 
	     &{\rm pressure\ max.} & & & & &{\rm missing} & {\rm def=0.4} & {\rm missing} & {\rm rate} \\
	     &({\rm Myr}) & &(10^8{\rm M}_{\odot}) &(10^8{\rm M}_{\odot}) &(10^8{\rm M}_{\odot}) & {\rm mass} &(10^8{\rm M}_{\odot}) & {\rm mass} & {\rm M}_{\odot}/{\rm yr} \\
	    \hline
	    {\rm NGC~4388} & 100^h &0.8^a & 3.4^b & 3.6^a & 23 & 18 & 9.2 & 61 & 5.6 \\
	    \hline
	    {\rm NGC~4402} & >300^i &0.5^c & 0.3^c & 4.4^c & 14 & 3 & 5.6 & 23 & <0.4 \\
	    \hline
	    {\rm NGC~4438} & 10^j &0.9^d & 5.0^e & 6.0^f & 48 & 12 & 19 & 38 & 130 \\
	    \hline
	    {\rm NGC~4501} & -100^k &0.5^a & <0.1^f & 17^a & 54 & <0.3 & 21 & 3 & - \\
	    \hline
	    {\rm NGC~4522} & 50^l &0.6^g & 1.7^g & 4.3^g & 17 & 14 & 6.8 & 68 & 5.0\\
	    \hline
	    {\rm NGC~4569} & 300^m & 1.2^a & 0.5^m & 6.0^m & 95 & 0.6 & 38 & 1.6 & 10.7\\
	    \hline
        \end{array}
      \]
\begin{list}{}{}
\item[$^a$] Cayatte et al. (1994)
\item[$^b$] Oosterloo \& van Gorkom (2005)
\item[$^c$] Crowl et al. (2005)
\item[$^d$] We took the HI deficiency of NGC~4579 which has about the same B and H band magnitudes and
morphological type. As NGC~4438, it also shows a highly truncated gas disk with a very small 
amount of H{\sc i}.
\item[$^e$] CO gas mass from Vollmer et al. (2006)
\item[$^f$] from this paper
\item[$^g$] Kenney et al. (2004)
\item[$^h$] Vollmer \& Huchtmeier (2003)
\item[$^i$] Based on the low mass and low surface brightness extraplanar H{\sc i}.
\item[$^j$] Vollmer et al. (2005)
\item[$^k$] Vollmer et al., in prep.
\item[$^l$] Vollmer et al. (2006)
\item[$^m$] Vollmer et al. (2004a)
\end{list}
\end{table*}

The percentage of extraplanar gas mass with respect to the expected gas mass
assuming an initially non-H{\sc i} deficient galaxy varies
between 0.3\% and 20\%. Thus, more than 80\% of the missing gas is undetectable in H{\sc i}.

\subsection{Initially gas deficient spiral galaxies}

The assumption of an initially non-H{\sc i} deficient galaxy is however questionable.
We argue that the interplay between ram pressure and evaporation of the galaxy's ISM
by the hot intracluster medium might be different between (i) the inner gas disk where star
formation occurs and the gas is clumpy and multiphase and (ii) the outer gas
disk where the atomic hydrogen is mainly warm ($T \sim 10^4$~K) and smoothly distributed
(see e.g. Braun 1997). In the inner star forming disk ($R < R_{25}$) the gas
is turbulent and clumpy giving rise to a tangled magnetic field which suppresses evaporation
(Cowie et al. 1981, Malyshkin \& Kulsrud 2001).
On the other hand, due to the kinematical quietness and the smoothness of the
outer gas disk, the magnetic field there is expected to be less tangled leading
to a much more efficient evaporation of the warm H{\sc i}.
Because of this effect together with the fact that the gas surface density decreases 
with increasing galactic radius in the outer gas disk, the outer gas is much more 
vulnerable to evaporation, harassment, turbulent/viscous stripping and ram pressure. 
We therefore argue that the outer gas disk is
stripped and evaporated at much larger distances from the cluster center than
one would expect from the Gunn \& Gott criterion (Gunn \& Gott 1972).
However, an early gas removal by ``preprocessing'' is also possible.

Based on this argument, we can recalculate the expected initial gas mass assuming
that the gas disk is truncated at the optical radius ($R_{25}$).
Spiral galaxies which show this property in the Cayatte et al. (1994) sample have
a mean H{\sc i} deficiency of 0.4. We thus recalculated the expected atomic gas mass
assuming this initial H{\sc i} deficiency (Table~\ref{tab:masses} col.(8)) and the
percentage of extraplanar gas mass with respect to the expected gas mass 
(Table~\ref{tab:masses} col.(9)). These percentage lie between 3\% and 70\%.
The galaxies that we observe close or up to 100~Myr after peak ram pressure
still have about half of the stripped gas in neutral form.

As a final step we can estimate an evaporation rate by 
\begin{equation}
\dot{M}_{\rm evap} \sim (M_{\rm HI}^{\rm def=0.4} - M_{\rm HI}^{\rm total})/t_{\rm rps}\ ,
\label{eq:evap}
\end{equation}
where $M_{\rm HI}^{\rm def=0.4}$ is the expected H{\sc i} mass assuming an initial
H{\sc i} deficiency of 0.4, $M_{\rm gas}^{\rm total}$ is the observed total H{\sc i} 
mass (from table~\ref{tab:masses}), and $t_{\rm rps}$ is the time to peak ram pressure.
This evaporation rate can be found in Table~\ref{tab:masses} col.(10).
We took here the time to peak ram pressure as the characteristic timescale.
Because it is not possible to determine if evaporation happened faster,
this timescale is an upper limit and, consequently, the derived evaporation
rate represents a lower limit.

\subsection{Evaporation rates}

Since NGC~4501 is in a pre-peak ram pressure phase, we cannot estimate
an evaporation timescale. It is intriguing that the three galaxies which are
only affected by ram pressure and which are observed less than 400~Myr after
peak ram pressure show the same evaporation rate of about 
$\dot{M}_{\rm evap} \sim 5-11$~M$_{\odot}$yr$^{-1}$.
The case of NGC~4438 is complicated, because of the unknown H{\sc i}
deficiency, the additional tidal and
ISM-ISM interactions, and a possible phase transition of the displaced gas,
but since ram pressure has the greatest effect on its ISM the derived
evaporation rate might still be valuable.

The analytical estimate of the classical evaporation rate of a spherical gas cloud 
by Cowie \& McKee (1977) is
\begin{equation}
\dot{M}_{\rm evap} = 4.34 \times 10^{-22} T_{\rm ICM}^{\frac{5}{2}}R_{\rm pc}(30/\ln \Lambda)\ {\rm M_{\odot}yr^{-1}}\ .
\end{equation}
For a cloud size of the order of the disk height $R_{\rm pc}=500$~pc, an ICM temperature
of $T_{\rm ICM}=3 \times 10^7$~K, and a Coulomb logarithm of $\ln \Lambda = 30$,
one finds $\dot{M}_{\rm evap} \sim 1$~M$_{\odot}$yr$^{-1}$.
About 1000 clouds are necessary to fill an H{\sc i} disk of constant hight $H=500$~pc between 
$R=5$~kpc and $R=10$~kpc. If these clouds have a density of $n = 1$~cm$^{-1}$ the resulting
gas mass is about $1.5 \times 10^9$~M$_{\odot}$. If 10\% of the surface of these clouds are
surrounded by the hot intracluster medium (this depends on the tail geometry)
the resulting total evaporation rate
is $\dot{M}_{\rm evap}^{\rm tot} \sim 100$~M$_{\odot}$yr$^{-1}$.

Our derived evaporation rate of NGC~4438 is close to that value. 
This might imply that the evaporation timescale is short (of the order of 10~Myr).
If we assume this short timescale for NGC~4388 and NGC~4522 we obtain evaporation
rate of 56~M$_{\odot}$yr$^{-1}$ and 25~M$_{\odot}$yr$^{-1}$ which are close to the
analytical estimate.

\subsection{Amount of detected stripped gas}

The percentage of the observed gas mass varies between 23\% and 68\% of the expected gas mass assuming an
initial H{\sc i} deficiency of 0.4 for the galaxies which are stripped inside the optical radius 
(table~\ref{tab:masses} col.(9)).
We argue that this corresponds to the fraction of cold H{\sc i} in the galactic disk before 
it was stripped. This is entirely consistent with the results based  on the deep H{\sc i}
observations of nearby undisturbed gas disks reported in (Braun 1997; see Sec.~\ref{sec:introduction}). 
We propose a scenario
where the diffuse warm ($T \sim 8000$~K) H{\sc i} evaporates rapidly and the cold
($T \sim 100$~K) H{\sc i} resists much longer and can still be observed 100~Myr
after its removal from the galactic disk. 
This scenario is also consistent with the H{\sc i} observations and simulations of NGC~4522
(Kenney et al. 2004, Vollmer et al. 2006) where a low surface brightness H{\sc i} component
is detected with a large linewidth ($\sim 100$~km\,s$^{-1}$). Vollmer et al. (2006)
interpreted this component as diffuse warm H{\sc i} which is stripped more efficiently
than the cold dense H{\sc i}. Here we add the effect of evaporation to this picture
which might be partly responsible for the increased stripping efficiency.

\subsection{The exception: NGC~4388 \label{sec:n4388}}

NGC~4388 is the only galaxy where extraplanar low surface density gas is detected
at distances larger than 10~kpc from the galactic disk. Oosterloo \& van Gorkom (2005)
speculated that the H{\sc i} tail did not evaporate, because NGC~4388 was stripped
by the ICM of M86. This ICM is less dense and maybe cooler than that of the Virgo
cluster, i.e. M87, and therefore evaporation is slower. However, M86 has a negative
absolute velocity and lies far behind M87 (more than 1~Mpc, see, e.g., Vollmer et al. 2004c).
On the other hand, NGC~4388 is located close to M87 ($D \sim 2000$~km\,s$^{-1} \times 100$~Myr
$\sim 0.1$~Mpc). The deprojected tail length in a putative M86 stripping scenario would be
$\sim 1$~Mpc which requires an unreasonable large evaporation time.
The most forward explanation, which is consistent with out findings, is that NGC~4388 is the only 
galaxy that we are observing in an evolutionary stage long enough after peak ram pressure
to show an extended gas tail and short enough after peak ram pressure so that the
tail is not yet evaporated. This time window has a width of $\sim 200$~Myr compared to
a cluster crossing time of $\sim 3$~Gyr explaining the low probability to observe a
galaxy in this evolutionary stage.

\section{Conclusions \label{sec:conclusions}}

We made deep H{\sc i} observations with the Effelsberg 100-m telescope around 5 H{\sc i} deficient
Virgo spiral galaxies to search for neutral gas located far away from the galactic disks
(more than 20~kpc). These galaxies are or were all affected by ram pressure stripping.
The following results were obtained
\begin{itemize}
\item
we did not detect H{\sc i} emission far away from NGC~4402, NGC~4438, NGC~4501, and NGC~4522;
\item
the already known H{\sc i} tail in the north of NGC~4388 does not extend further than
the WSRT image of Oosterloo \& van Gorkom (2005) has shown;
\item
the H{\sc i} tail of NGC~4388 seems thus to be an exception.
\end{itemize}
Based on the absence of H{\sc i} tails in these galaxies and a balance of previous
detections of extraplanar gas in the targeted galaxies we propose a global picture
where the outer gas disk (beyond the optical radius $R_{25}$) is evaporated/stripped 
much earlier than expected by the classical ram pressure criterion (Gunn \& Gott 1972).
The key ingredient for this argument is the two-phase nature of the atomic hydrogen.
In the inner disk ($R<R_{25}$) cold and warm H{\sc i} coexist, whereas in the outer
disk the atomic gas is mostly warm (Braun 1997). The cold H{\sc i} is located
near star forming regions and might be stirred by supernova explosions.
This dynamical stirring causes a tangled magnetic field in the dense cold H{\sc i}
clouds which inhibits their evaporation by the hot intracluster medium once they
are pushed out of the galactic disk by ram pressure. We further argue that the
warm diffuse H{\sc i} is evaporated and stripped rapidly with an evaporation
rate between 10 and 100~M$_{\odot}$yr$^{-1}$.
After a ram pressure stripping event we therefore can only observe the fraction of the 
ISM which was in form of dense cold clouds before it was removed from the galactic disk.
More observations are needed to test our scenario of the stripping of a two-phase atomic
hydrogen.

\begin{acknowledgements}
Based on observations with the 100-m telescope of the MPIfR (Max-Planck-Institut f\"{u}r 
Radioastronomie) at Effelsberg.
\end{acknowledgements}

\end{document}